\documentclass[aps,final,notitlepage,oneside,twocolumn,nobibnotes,nofootinbib,
superscriptaddress,noshowpacs,centertags,showkeys]{revtex4-1}

\usepackage[utf8]{inputenc}
\usepackage[english]{babel}
\usepackage{graphicx}
\usepackage{latexsym}
\usepackage{amssymb}
\usepackage{amsmath}
\usepackage{float}
\usepackage{color}

\begin{document}

\title{Strong gravitational lensing of blazar gamma-radiation and intergalactic magnetic fields}
\author{Yu.N. Eroshenko}\thanks{e-mail: eroshenko@inr.ac.ru}
\affiliation{Institute for Nuclear Research of the Russian Academy of Sciences
60th October Anniversary Prospect 7a, 117312 Moscow, Russia}

\date{\today}

\begin{abstract}
The influence of intergalactic magnetic fields on the strong gravitational lensing of blazar secondary gamma radiation is  discussed. Currently, two cases of strong gravitational lensing of blazar gamma-radiation are known, where radiation is deflected by galaxies on the line of sight between the blazars and the Earth. The magnetic field can affect the movements of electron-positron pairs generated by primary radiation, and thereby change the directions of secondary gamma radiation. It modifies the equation of the gravitational lens and leads to the dependence of the observed signal in the secondary gamma radiation on the energy of photons and on the magnetic field. Accordingly, it is possible in principle to estimate the intergalactic magnetic fields from the time delay of signals, from the angular position of images (for future high-resolution gamma-ray telescopes) or from the shape of the observed energy spectrum. This method is demonstrated by the example of the blazar B0218+357. In this case however it is not possible to obtain useful constraints due to the large distances to the blazar and the lens galaxy. The result is only a lower limit on the magnetic field $B>2\times10^{-17}$~G, which is weaker than other existing constraints. But the future discoveries of lensed blazars may provide more favourable opportunities for measuring the magnetic fields, especially with the help of new generation of gamma-ray telescopes such as e-ASTROGAM, GECAM, and SVOM as well as future gamma-ray telescopes with high angular resolution, $\sim0.1''$. 
\end{abstract}

\maketitle


\section{Introduction}
\label{introsec}

The origin of intergalactic magnetic fields has not yet been reliably clarified, although some models of their generation have been proposed \cite{DurNer13}. Probably, early seed fields appeared first (including at the stage of inflation) or these seed fields were born in the first stars or protogalaxies. Later, the fields could be amplified by the dynamo effect during the large-scale structure formation or in stars. Typical measured magnetic field $B$ is of the order of tens of $\mu$G (in galaxies) and of the order of $\mu$G (in galaxy clusters), however, for fields outside galaxy clusters and in the voids, there are only restrictions $10^{-16}$~G$\leq B\leq10^{-9}$~G, where the lower value is derived from the unobservability of the gamma halo \cite{DurNer13}.

In the work \cite{NerSem07} (see also \cite{ElyNerSem09,Doletal09}), a method for measuring magnetic fields in voids from the angular profiles of the ``gamma halos'' around the images of blazars was proposed. Blazars are active galactic nuclei, whose relativistic jets are directed at small angles to the direction of the Earth. X-rays, primary gamma radiation and possibly cosmic rays are emitted along the collimated jets. Photons of primary gamma radiation, when interacting with intergalactic background radiation, give rise to electron-positron pairs \cite{Nik62}, these pairs move and deflected in intergalactic magnetic fields for some time and then produce secondary gamma radiation in the process of reverse Compton scattering. Due to the deflection of $e^+e^-$, the images of blazars in secondary gamma rays turns out to be blurred, and from the angular structure of the gamma halos, it will be possible to determine the magnetic field or constrain it. This method is applicable for measuring fields in the range $10^{-16}$~G$<B<10^{-12}$~G. An additional possibility is to observe the time delay between the arrival of primary and secondary radiation in the presence of flash activity \cite{Pla95}. The work \cite{AndKus10} reported the possible detection of a gamma halo whose properties are compatible with $B\sim10^{-15}$~G, but independent studies have not yet confirmed this result \cite{Neretal11,Acketal13}. Other indications of the gamma halo were obtained in \cite{CheBucFer14}.

The formation of cascades in interaction with intergalactic background radiation also modifies the gamma-ray spectrum, reducing the flux \cite{Dogetal11}, \cite{FinReyGeo13}. Comparison of the blazar spectrum in the high-energy part recorded by ground-based Cherenkov detectors and in the low-energy part available with Fermi-LAT allowed one to conclude that intergalactic magnetic fields $B>10^{-19}$~G at a confidence level $>5\sigma$ if the coherence length of the magnetic field $l_c\sim B/|\nabla{\bf B}|\geq1$~Mpc \cite{Finetal15}. However, this restriction is weaker than the above one, $B\geq10^{-16}$~G.

Note that the oscillation of gamma photons into axion-like particles in magnetic fields was also considered  by \cite{MeyMonCon14,Tro15}. This process could explain the transparency of the Universe in gamma rays at high energies, which is required to explain the observations of some distant blazars. However, these observations can also be explained by the generation of gamma radiation in cascade processes \cite{BeretalCascads} during the propagation of ultrahigh-energy cosmic rays, if the blazars are sources of such cosmic rays \cite{InoKalKus14}. 

In this paper, we propose a new method for studying intergalactic magnetic fields based on observations of strong gravitational lensing of blazar secondary gamma radiation. Strong gravitational lensing leads to the appearance of multiple images of an object (weak lensing distorts only its shape) when light is deflected by the gravitational field of the lens galaxy located on the line of sight. Strong gravitational lensing is a powerful tool for exploring the early universe. For example, recently this effect helped to detect and confirm spectroscopically the population of galaxies responsible for the process of reionization of the Universe to redshifts $z\sim6$ \cite{Ateetal23}. To date, more than 200 examples of strong gravitational lensing of quasars are known in the optical and radio ranges, there are two cases of strong gravitational lensing of gamma-ray blazars PKS~1830-211 \cite{BarGliMou11,Baretal15,Abdetal15} and B2~0218+357 \cite{Cheetal14}, and such observations should become more numerous. According to the calculations of \cite{BarBotSus14}, for about 30\% of blazars, there should be a galaxy on the line of sight that creates strong gravitational lensing.  Observations in the optical and radio ranges allow measuring distances with good accuracy and determining the mass of the lens, i.e. restoring the entire lensing configuration. Thus, when studying the lensing of gamma radiation, this configuration will be known from observations in other ranges.

The proposed method for studying intergalactic magnetic fields is based on the fact that the magnetic field changes the direction of motion of electron-positron pairs generated by primary radiation. For this reason, secondary gamma radiation is emitted at different angles with respect to the primary radiation. Such a change in direction modifies the equation of the gravitational lens. In this case, there is a dependence of the lens amplification on the radiation energy and on the magnetic field, while conventional gravitational lensing occurs achromatically. The deflection of charged particles (cosmic rays) in cosmic magnetic fields is sometimes called magnetic lensing (see \cite{HarMolRou99,Haretal02,KacSerTes05,DolKacSem09,BatCasMas11}). The effect we are considering can be called a combination of gravitational and magnetic lensing. Due to the influence of the magnetic field, the energy spectrum of secondary gamma radiation is modified (the amplification is not achromatic), and from the spectrum, information can be obtained about the intergalactic magnetic fields on the way of electron-positron pairs. 

The angular resolution of gamma-ray telescopes is still small, and individual images in gamma rays due to strong gravitational lensing at cosmological distances are unlikely to be resolved in the near future. However, if the future gamma-ray telescopes will be able to resolve individual images, then it will be possible to measure magnetic fields from the angular distribution of lensed radiation. 

Another possible effect, which can be observed only with sufficiently weak magnetic fields, is the time delays between images during blazar flash activity. Secondary gamma radiation is usually averaged over long time intervals, since $e^+e^-$-pairs fly far away from the line of sight to the blazar, and only in the case of small magnetic fields is will be possible to observe flash activity in the secondary radiation at times acceptable for observations. When observing the lensing of the gamma radiation of the blazar PKS~1830-211 at a redshift of $z=0.89$ \cite{BarGliMou11}, the spatial resolution of the Fermi LAT gamma telescope was insufficient to observe individual images, but a correlation with a time shift of $27.5\pm 1.3$ days was found in the variable signal, corresponding to the time delay between images measured in the radio range. Similar studies of the time delay in gamma radiation were performed for the blazar B0218+357 \cite{Cheetal14,Sitetal15}. In the work \cite{Cheetal14} it was found that the delay between the lensed gamma signals is $11.46\pm0.16$ days, which is $\sim1$ day more than the delay between radio signals. In this paper we will show that such a difference can be explained by the influence of the intergalactic magnetic field. The time delay effect appears to be the promising method of measuring magnetic fields from the gravitational lensing, and we demonstrate its application in one model example. 

The strong gravitational lensing of gamma-ray blazars has already been considered in other works in some aspects unrelated to intergalactic magnetic fields. For example, in the work \cite{Baretal14}, the observation of a relativistic blazar jet through a gravitational lens was considered as a promising method for studying the structure of the jet, and for localizing the place of gamma radiation generation along the jet. In the work \cite{BarBotSus14}, the influence of the gas halo and the radiation field of the lens galaxy on the gamma radiation passing near it was considered, however this effect is usually insignificant because gamma radiation passes far from the lens galaxy. 

The article is structured as follows. Section~\ref{prisecsec} provides some basic formulas necessary for further calculations. In the section~\ref{lenzsec}, the equation of the gravitational lens is derived and solved, taking into account the additional segments. In the section ~\ref{dist}, the distance between the birth birth places of the $e^+e^-$ pairs is calculated. In the section~\ref{delay}, the time delay between the lensed images is obtained depending on the magnetic field. In the section~\ref{spectrumsec}, the transformation of the gamma-ray spectrum of blazars under the influence of a gravitational lens is discussed and a method for determining intergalactic magnetic fields from this effect is proposed. The section~\ref{twosec} discusses the option when two images created by a gravitational lens can be resolved. Finally, in the section~\ref{conclsec} we briefly discuss the results obtained.


\section{Primary and secondary gamma radiation}
\label{prisecsec}

The extragalactic background radiation, composed by IR and UV photons $\gamma_{\rm EBL}$ accumulated over
cosmological time from the radiation of stars and from the re-emission by dust.
The free path length of $\gamma$-photons with energy $E_{\gamma0}$ relative to the process $\gamma+\gamma_{\rm EBL}\to e^+e^-$ is \cite{DurNer13}
\begin{equation}
D_{\gamma 0}\simeq0.8\left(\frac{E_{\gamma0}}{1\mbox{~TeV}}\right)^{-1}~\mbox{Gpc},
\label{dg0eq}
\end{equation}
and the energy of the electron or positron being born is $E_e\simeq E_{\gamma0}/2$ with the threshold $E_{\gamma0}\simeq250$~GeV.

The $e^+$ or $e^-$ with initial energy $E_e$ loses energy due to inverse Compton scattering by photons of relic radiation at the distance \cite{DurNer13}
\begin{equation}
\lambda_e=\frac{3m_e^2}{4\sigma_T\rho_r E_e}\simeq0.233\left(\frac{E_{\gamma0}}{1\mbox{~TeV}}\right)^{-1}~\mbox{Mpc},
\label{ic3}
\end{equation}
where $m_e$ is the mass of the electron,  $\sigma_T$ is the Compton cross-section, and $\rho_r$ is the density of relic radiation. This proses is accompanied by the emission of the secondary photons, and the relationship between the energy of the primary photon $E_{\gamma0}$ and the energy of the secondary photon $E_{\gamma}$ has the form 
\begin{equation}
E_\gamma=\frac{4E_rE_e^2}{3m_e^2}\simeq0.75\left(\frac{E_{\gamma0}}{1\mbox{~TeV}}\right)^2~\mbox{GeV},
\label{relation}
\end{equation}
where the mean energy of the relic radiation photon is $E_r=6\times10^{-4}$~eV.

The Larmor radius of the $e^\pm$  trajectory is
\begin{equation}
r_L=\frac{E_e}{eB}\simeq5.4\times10^{-2}\left(\frac{E_{\gamma0}}{1\mbox{~TeV}}\right)\left(\frac{B}{10^{-14}\mbox{~G}}\right)^{-1}~\mbox{Mpc},
\end{equation}
and the deflection angle of an electron or positron in a magnetic field \cite{NerSem07} 
\begin{equation}
\delta\simeq\frac{\lambda_e}{r_L}\simeq4.31\left(\frac{E_{\gamma0}}{1\mbox{~TeV}}\right)^{-2}\left(\frac{B}{10^{-14}\mbox{~G}}\right)\mbox{~rad},
\label{deltaeq}
\end{equation}
where the expression (\ref{ic3}) for $\lambda_e$ was accepted here. It should be noted, however, that the electron emits many photons and an electromagnetic cascade develops. Losing energy, the electron is deflected at larger angles. Therefore, the expression (\ref{deltaeq}) refers only to the emission of the first most energetic photon and has the meaning of the minimum deflection angle. If $\lambda_e>l_c$, then the deviation of the trajectory has a diffusive character, and the angle $\delta$ can not be obtained from (\ref{deltaeq}). To illustrate the method, we consider only the most energetic secondary photons that deviate by minimal angles.


\section{The equation of the lens with intermediate $e^+e^-$ pairs}
\label{lenzsec}

The theory of gravitational lensing is presented, for example, in \cite{MolRou02,SchEhlFal92}, where, in particular, the equation of the gravitational lens is derived. In this section, we obtain a similar equation for the case when, at a distance of $D_{\gamma 0}$ from the source, a gamma photon generates $e^+e^-$ pairs. The electrons (positrons) of the pairs are deflected in the magnetic field by an angle of $\delta$ and generate secondary gamma photons in the process of reverse Compton scattering. While the low-energy part of the primary gamma radiation is lensed in the same way as the optical and gamma radiation of the blazar, the photons of the secondary gamma radiation follow different trajectories, because the $e^+$ and $e^-$ producing them have deflected in the intergalactic magnetic field. 

\begin{figure}[t]
\begin{center}
\includegraphics[angle=0,width=0.48\textwidth]{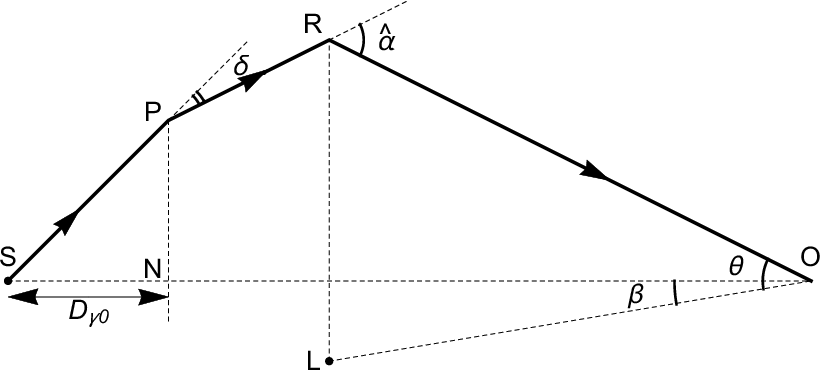}
\end{center}
\caption{The configuration of gravitational lensing taking into account the rotation of the trajectory at point P by an angle of $\delta$ due to the deflection of $e ^\pm$ in magnetic field. S is a blazar (source of primary gamma radiation), L is a gravitational lens, O is an observer registering secondary gamma radiation. The relative vertical scales are increased, though in reality all the angles shown are small.  \label{gr1}}
\end{figure}

Suppose first that the distance $D_{\gamma 0}$ is less than the distance from the source to the lens $D_{LS}$, so that the lens deflects the secondary photons. Thus, at the distance $D_{\gamma 0}$ from the source, there is a break of the beam trajectory at an angle of $\delta$ (see Fig.~\ref{gr1}), which must be taken into account in the lens equation. 

The angle $\hat\alpha$ (all angles are assumed to be small) is the deflection of light passing at a distance $r_m$ from the point mass $M$,
\begin{equation}
\hat\alpha=\frac{4GM}{c^2r_m},
\end{equation}
the and Einstein angular radius 
\begin{eqnarray}
\theta_E&=&\sqrt{\frac{D_{LS}}{D_{OS}D_{OL}}\frac{4GM}{c^2}} 
\label{newlab1} 
\\
&=&1.4\times10^{-5}\left(\frac{1\mbox{~Gpc}\times D_{LS}}{D_{OS}D_{OL}}\right)^{1/2}\left(\frac{M}{10^{12}M_\odot}\right)^{1/2}.
\nonumber 
\end{eqnarray}
Then the reduced deflection angle
\begin{equation}
\alpha=\hat\alpha\frac{D_{LS}}{D_{OS}}=\frac{\theta_E^2}{\theta}.
\end{equation}
The equation of an ordinary gravitational lens is written as 
\begin{equation}
\theta=\beta+\alpha.
\label{lenseq1}
\end{equation}

In our case, the geometry of the path of the light beam includes an additional rotation by the angle $\delta$ (see Fig.~\ref{gr1}). Therefore, the lens equation takes the following form
\begin{equation}
\theta=\beta+\alpha+\delta\frac{D_{\gamma 0}}{D_{OS}}
\label{lensini2}
\end{equation}
Denote for brevity $\tilde\delta=\delta D_{\gamma 0}/D_{OS}$ and rewrite the equation of the lens as
\begin{equation}
\theta^2-(\beta+\tilde\delta)\theta-\theta_E^2=0
\label{lenseq2}
\end{equation}
Its solutions ($\pm$ correspond to two images) 
\begin{equation}
\theta_\pm=\frac{\beta+\tilde\delta}{2}\pm\theta_E\sqrt{1+\frac{(\beta+\tilde\delta)^2}{4\theta_E^2}}
\label{sol}
\end{equation}
in the limit $\tilde\delta=0$ are reduced to the known solutions of the equation (\ref{lenseq1}) given in \cite{MolRou02}. If the objects are located at cosmological distances, then the distances in the above formulas should be understood as cosmological distances that determine the angles \cite{MolRou02}.

Note that (\ref{sol}) includes the sum of $\beta+\tilde\delta$, so the value $\tilde\delta$ plays the role of an additional angular shift of the gravitational lens from the $SO$ line of sight. The magnetic field plays a decisive role in the deflection of the beam when
$\tilde\delta\geq\beta$. According to (\ref{deltaeq}), this condition for the first photon emitted in the process of reverse Compton scattering has the form
\begin{equation}
E_{\gamma0}\leq7\left(\frac{\beta}{0.01}\right)^{-1/3}
\left(\frac{B}{10^{-14}\mbox{~G}}
\right)^{1/3}\left(\frac{D_{OS}}{1\mbox{~Gpc}}
\right)^{-1/3}\mbox{~TeV}.
\label{e7}
\end{equation}
At high energies, two images of the blazar will be observed in gamma rays, at the same angles as in optical observations. If (\ref{e7}) is performed, then there will actually be no gravitational lensing. The observed image of the blazar will be surrounded by a gamma halo in this case, as shown in \cite{NerSem07}, and the gravitational lens will distort only a small portion of this halo.

An electron or positron loses energy as photons are emitted and deviates at even greater angles, so the subsequent emitted photons also satisfy the criterion $\tilde\delta\geq\beta$. If the initial photon has $E_{\gamma0}$ greater than (\ref{e7}), then the initial emitted photons do not satisfy the criterion $\tilde\delta\geq\beta$, but the subsequent ones can satisfy. Note that among all the secondary photons, only those photons that were born in the planes passing through the Earth and the lens galaxy reach the observer along the optical path of the lens. At the same time, the blazar itself (the source of primary photons) is no longer obliged to lie in this plane, as it would be in the case of conventional gravitational lensing. Accordingly, when secondary gamma radiation is lensed, the two images will be in the form of two arcs with angular size $\theta_\pm$. Electrons and positrons are deflected in opposite directions by angles $\pm\delta$, so each of the two images splits into two more parts.  

The opening angle of the blazar jet has some finite value $\theta_{jet}\sim3^\circ-5^\circ$, therefore, depending on the direction of the jet, only one of the trajectories (primary gamma radiation) of the gravitational lens can get inside the jet, or all the rays will pass outside the jet, and then the gravitational lensing will not happen.

Note that in the case of weak magnetic fields, the modified equation of the gravitational lens (\ref{lensini2}) cannot be interpreted as the gravitational lensing of some extended object (cloud $e^\pm$) arising at a distance of $D_{\gamma 0}$ from the blazar. This is due to the fact that the direction of the $e^\pm$ movement and, accordingly, the direction of the secondary photons are important. 

Let us now consider the case $D_{\gamma 0}>D_{LS}$, when the gravitational lens deflects the primary gamma radiation. The geometry in this case leads to the same equation of the lens (\ref{lensini2}). Note, however, that in order to fulfil the condition $D_{\gamma 0}>D_{LS}$, the lens galaxy should be located very close to the source, and such a configuration is very rare, therefore we do not consider this case further.


\section{The distance between $e^+$ and $e^-$ birth regions and the coherence length of the magnetic field}
\label{dist}

It should also be taken into account that $e^+$ and $e^-$ in each of the two optical paths are deflected by the magnetic field in different directions, so there are four different paths of secondary photons. At the same time, the magnetic field in each of the four regions may be different in magnitude and direction if these regions are separated by a distance exceeding the magnetic field coherence length $l_c$. At the $l_c$ scale, the magnetic field changes by the amount of $\Delta B\sim B$, so the angle of deflection also varies greatly $\Delta\delta\sim\delta$. Moreover, even if the specified regions in each of the ``+'' and ``--'' paths are located closer than $l_c$, the distance between ``+'' and ``--'' pairs may exceed $l_c$. Thus, the problem arises of considering various cases with a different ratio of distances between the  $e^+$ and $e^-$ birth places and $l_c$, and different variants of random variables $B$ in the case when the birth places are spaced further than $l_c$.  

Let us find the distance $L$ (the length of the segment PN in Fig.~\ref{gr1}) of the $e^+e^-$ birth places from the line of sight $SO$. From the geometry of ray propagation one has
\begin{equation}
L_\pm=D_{\gamma 0}(\hat\alpha+\beta-\theta_\pm+\delta).
\label{lini}
\end{equation}
Depending on the numerical values, $L_\pm$ can either exceed the coherence length $l_c$ or be less than it. 
The distance between the $e^+$ and $e^-$ birth places  in one path ``+'' is
\begin{equation}
\Delta L_+=L_+^{(1)}-L_+^{(2)},
\label{dlplus}
\end{equation}
where $e^+$ is denoted by the index ``(1)'', and $e^-$ by the index ``(2)''. In the case of $D_{\gamma 0}\ll D_{OS}$, taking into account (\ref{lini}), the value (\ref{dlplus}) can be estimated as 
\begin{equation}
\Delta L_+\sim 2\delta D_{\gamma 0}=0.6\left(\frac{E_{\gamma0}}{5\mbox{~TeV}}\right)^{-3}\left(\frac{B}{10^{-14}\mbox{~G}}\right)\mbox{~Mpc}.
\label{dlplus2}
\end{equation}
If $B\gg10^{-14}$~G, then in most cases, at the energies we are considering, the value (\ref{dlplus2}), and even more so (\ref{lini}), exceeds 1~Mpc.


\section{Time delay}
\label{delay}  

Let's consider the question of the time delay between the signals that have propagated in two ways around a gravitational lens. Calculating the geometric delay (between the cases of the presence and absence of a lens) in the presence of an additional deviation by the angle $\delta$ instead of the Eq.~(3.36) in \cite{MolRou02} now gives
\begin{equation}
\delta t_{geom}=\frac{D_{OS}D_{OL}}{2cD_{LS}}\left[(\theta-\beta)^2-\tilde\delta(\theta-\beta)\right].
\end{equation}
It should be noted that for $e^+$ and $e^-$, the angle $\tilde\delta$ has opposite signs, so there will be four components in the received signal that come earlier and later than in case of the $e^\pm$ absence. 

The gravitational time delay (Shapiro delay) was calculated, e.g., in \cite{Wei72}
\begin{equation}
\delta t_{grav}=\frac{2GM}{c^3}\left(\frac{r-r_0}{r+r_0}\right)^{1/2}+\frac{2GM}{c^3}\ln\left(\frac{r+\sqrt{r^2-r_0^2}}{r_0}\right),
\end{equation}
where, in the point lens approximation, $r_0$ is the minimum distance of the trajectory to the lens, $r$ is the distance of observer from the lens, $M$ is the mass of the lens galaxy, and $G$ is the gravitational constant. In our case, one can evaluate
$r_0/r\simeq D_{OL}\theta$. The full time delay $\delta t(\theta,\tilde\delta)=\delta t_{geom}+\delta t_{grav}$ depends on the $\theta$ ($\theta=\theta_\pm$for two paths) and $\tilde\delta$.

Consider first the case when $L_++L_-<l_c$, i.e. the magnetic field for all four regions of $e^+$ and $e^-$ is the same. The difference in the signal travel time along two paths 
\begin{equation}
\Delta t(\tilde\delta)=\delta t(\theta_+,\tilde\delta)-\delta t(\theta_-,\tilde\delta)
\end{equation} 
With the above equations one has
\begin{eqnarray}
\Delta t(\tilde\delta)&=&\frac{2GM}{c^3}\ln\left|\frac{\theta_+}{\theta_-}\right|+
\label{dtfin}
\\
&+&\theta_E\sqrt{1+\frac{(\beta+\tilde\delta)^2}{4\theta_E^2}}\left(\frac{4GM}{c^3}-\frac{D_{OS}D_{OL}}{2cD_{LS}}\beta\right).
\nonumber
\end{eqnarray}

When observing the lensing of the blazar in the radio range, there will be no correction from $\tilde\delta$, therefore, the relative shift of the difference in the course of the rays in the gamma and in the radio range is
\begin{equation}
\Delta=\frac{\Delta t^{\gamma}-\Delta t^{radio}}{\Delta t^{radio}}=\frac{\Delta t(\tilde\delta)-\Delta t(\tilde\delta=0)}{\Delta t(\tilde\delta=0)}.
\label{deleq}
\end{equation}
In the case of the blazar B0218+357 discussed below, the gravitational time delay turns out to be more important than the geometric one.

If $\tilde\delta\sim\beta$, then the shift of the time arrival difference of the signals can be comparable to the time difference itself $\Delta\sim1$. Moreover, in the case of $\tilde\delta\leq\beta$, the signal in the secondary gamma radiation will not be averaged over long periods, and flash activity in the radio range and at high energies can 
can be compared in the same case of gravitational lensing. Then, based on the delay of the signals, it will be possible to draw a conclusion about intergalactic magnetic fields. This method differs from the method proposed in \cite{Pla95} for searching for echo signals between primary and secondary gamma radiation from blazars without gravitational lensing. The problem can only be to separate the secondary gamma radiation against the background of the primary one, which can also be variable. Such separation can be performed according to the Fourier analysis of time profiles. The primary gamma radiation will be lensed in the same way as the radio emission of the blazar, and the secondary will have a different time delay depending on the energy. 

More precisely, we should not talk about the delay of the signal, but about the broadening or splitting of the lensed pulse into two parts, because $\tilde\delta$ has two opposite sign values for for $e^+$ and $e^-$. However, one of the images is usually more strongly enhanced by the gravitational lens and will prevail, so the broadening of the pulse will be asymmetric and will look like an additional time shift. In the case when $l_c$ is less than the distance between the birth regions of $e^+$ and $e^-$, several different options are possible, as explained earlier. 

We apply the formalism developed in this Section to the case of gamma-lensing of the B0218+357 blazar. In the work \cite{Cheetal14} it was found that the delay between the lensed gamma signals is $11.46\pm0.16$ days, which is $\sim1$ day more than the delay between radio signals. Note, however, that in \cite{BigBro18} a different result was obtained with less or no delay. If the difference is $\sim1$ day between the delay times really takes place and is due to the birth of $e^\pm$, then $\Delta\sim0.1$. The configuration of gravitational lensing was discussed in \cite{WucBigBro04,Yoretal05,Baretal15-2}. The blazar B0218+357 is at $z=0.944\pm0.002$, and the lens is at $z=0.6847$. It follows that $D_{LO}\approx1.48$~Gpc and $D_{SO}\approx1.65$~Gpc. The Einstein radius is $\theta_E\simeq0.1672\pm0.0006"$. The angular distance between the images is $0.3344"$. From here we obtain $\beta\simeq0.0035"$, and from numerical solution of (\ref{deleq}) we get $\tilde\delta\simeq0.00034"$. With this value, according to the formulas of the Section~\ref{prisecsec},
\begin{equation}
B>2\times10^{-17}\left(\frac{E_{\gamma}}{100\mbox{~GeV}}\right)^{3/2}\mbox{~G},
\label{b14eg}
\end{equation}
where $E_{\gamma}$ is the observed energy. Since Fermi-LAT performed observations at energies $E_{\gamma}>100$~MeV, we get that $B>2\times10^{-20}$~G. The MAGIC telescope system observed lensing events at $E_{\gamma}>$100~GeV \cite{Sitetal15}. According to (\ref{b14eg}), this gives $B>2\times10^{-17}$~G, which is comparable in order of magnitude to the constraints resulting from the absence of a gamma halo \cite{DurNer13}. Therefore, in the case of B0218+357, it is not possible to obtain useful constraints due to the large distances to the blazar and the lens galaxy. The observation of other cases of lensing may provide more favorable opportunities to find or constraint $B$.

Note that in the energy regions under consideration there must be a contribution from both primary and secondary gamma radiation, the separation of which is a difficult task. Secondary radiation can make a noticeable contribution to the delayed signal if the flare activity of the blazar occurs mainly at the highest energies of $\sim1-100$~TeV, since photons with such energies can produce cascades and secondary photons available for observations. If the primary radiation is also highly variable, then the flashes in the secondary radiation will be blurred, and it will be difficult to notice the specified shift by $\sim1$ day.


\section{The effect of a gravitational lens on the blazar gamma-ray spectrum}
   \label{spectrumsec}

Let's fix a small energy interval from $E_{\gamma0}$ to $E_{\gamma0}+dE_{\gamma0}$.
The angle $\delta$ in Fig.~\ref{gr1} depends on $E_{\gamma0}$, so the interval $E_{\gamma0}$ to $E_{\gamma0}+dE_{\gamma0}$ corresponds to a certain interval of angles from $\psi$ to $\psi+d\psi$. The amount of energy recorded in this interval depends on the solid angle corresponding to the interval from $\psi$ to $\psi+d\psi$. At the same time, it is necessary to recalculate the energy of the observed secondary photons according to the Eq.~(\ref{relation}).

From the geometry of the beam we obtain
\begin{equation}
\psi=\hat\alpha+\beta+\delta-\theta.
\end{equation}
If $D_{\gamma 0}=\lambda/E$ and $\delta=\varkappa B/E^2$, according to (\ref{dg0eq}) and (\ref{deltaeq}), then the differential amount of the observed energy is proportional to the following value
\begin{eqnarray}
&&\frac{d\psi_\pm}{dE_{\gamma0}}=\frac{\varkappa B}{E_{\gamma0}^3}\times
\nonumber
\\
&&\times\left[\frac{3\lambda}{2E_{\gamma0}D_{OS}}\left(
1+\frac{4\theta_E^2D_{OS}}{D_{LS}\left(\beta+\tilde\delta\pm\sqrt{4\theta_E^2+(\beta+\tilde\delta)^2}\right)}
\right)\right. \nonumber
\\
&&\left.\times \left(1\pm\frac{\beta+\tilde\delta}{\sqrt{4\theta_E^2+(\beta+\tilde\delta)^2}}\right)-2\right].
\label{dpsipm}
\end{eqnarray}
In the absence of a gravitational lens, only the effect of the cascade gamma radiation gives 
\begin{eqnarray}
\frac{d\psi}{dE_{\gamma0}}=\frac{\varkappa B}{E_{\gamma0}^3}\left[\frac{3\lambda}{E_{\gamma0}D_{OS}}-2\right].
\label{bezlinzy}
\end{eqnarray}
If the initial blazar spectrum has a power-law form 
\begin{equation}
\Phi(E)=\frac{dN}{dE}\propto E^{-\Gamma}, 
\end{equation}
where usually $\Gamma\simeq1.5-1.8$ \cite{Finetal15}, then after gravitational lensing it is converted to $\propto(d\psi/dE)E^{-\Gamma}$. Based on this change in the spectrum compared to the typical spectra of blazars, it could be concluded that the magnetic field played a significant role in the formation of the lensed image. Thus, in the presence of a gravitational lens, an additional dependence of the recorded amount of secondary gamma radiation energy on the magnetic field $B$ and the photon energy $E_{\gamma0}$ appears (through $\tilde\delta$ dependence). 

Denote the product of parentheses in (\ref{dpsipm}) by $S_\pm^i$, where the indices $i=1$ and $i=2$ denote, respectively, electrons and positrons. Then it can be shown that
\begin{equation}
S_+^1+S_-^1=S_+^2+S_-^2=2+\frac{D_{OS}}{2D_{LS}}=const,
\label{ssss}
\end{equation}
i.e. in the total signal (if four beams are not resolved by angles), the dependence on energy will not differ much from (\ref{bezlinzy}). Note, however, that the opening angle of the blazar jet has a finite value. Therefore, if the optical paths fall into the jet only on one side of the lens, then there will be no compensation like (\ref{ssss}), and the values of $S_+^1+S_+^2$ or $S_-^1+S_-^2$ will depend on the energy. Unfortunately, this effect is very small, because the corrections in (\ref{dpsipm}) compared to (\ref{bezlinzy}) in typical cases are only $\sim3$~\%, so one can see only the formation of a gamma halo around the blazar image and the transformation of the spectrum by the Eq.~(\ref{bezlinzy}), as indicated in \cite{NerSem07}. 

\begin{figure}[t]
\begin{center}
\includegraphics[angle=0,width=0.48\textwidth]{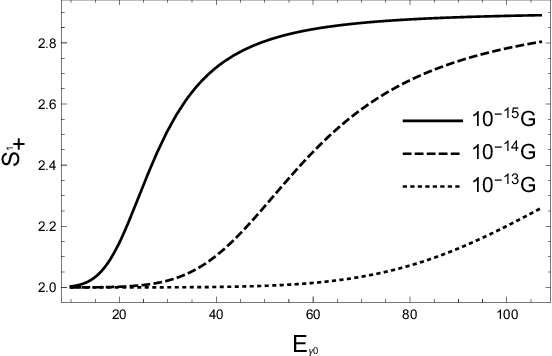}
\end{center}
\caption{Function $S_+^1$ describing the transformation of the blazar spectrum. The energy of the primary photons $E_{\gamma0}$ is given in TeV. The graph is given for parameter values: $D_{OS}/D_{LS}=2$, $\beta/\theta_E=0.2$, $D_{OS}=1$~Gpc, and for 3 values of magnetic field $B$.
\label{gr2}}
\end{figure}

The best prospects for the selection of the spectrum transformations caused by the lens will appear if it is possible to resolve individual images $S_+^1$, $S_+^2$, $S_-^1$, $S_-^2$. Although the compensation (\ref{ssss}) occurs in the sum of the images, in individual images the additional dependence on energy is quite strong. For example, in Fig.~\ref{gr2} shows the value of $S_+^2$ for some typical parameter values.


\section{The case when two images are resolved}
   \label{twosec}

If a gamma-ray telescope allows you to measure the angular distance between two images $\theta_+-\theta_-$ depending on the energy, then by fitting the observed theoretical dependence, one can find $B$. Using (\ref{sol}), we get
\begin{equation}
\theta_+-\theta_{-}=\sqrt{4\theta_E^2+(\beta+\tilde\delta)^2}.
\label{separate}
\end{equation}
Since the configuration of the gravitational lens will be known from optical and radio observations, the $\beta$ and other characteristics of the lens in (\ref{separate}) will be fixed. Therefore, if fitting allows one to find $\tilde\delta$, then from the dependence of $\tilde\delta$ on $B$ given by the Eq.~(\ref{deltaeq}) at different photon energies, one can find $B$. The application of this method is possible only in the future, when the angular resolution of gamma-ray telescopes will increase by several orders of magnitude, to the level of $\sim1"$.

If the intergalactic magnetic fields are large enough, then $e^+e^-$-pairs have time to deviate at large angles, and their trajectories become entangled. As a result, an extended region appears around the blazar, from which secondary photons are emitted. This region is called pair halo. It emits gamma photons as an extended source. In this case, the gravitational lens can create multiple images of the pair halo and amplify the gamma radiation flux in the same way as it amplifies the optical and radio radiation flux. With a sufficiently high resolution in the gamma range, an observer could see that the optical images of the blazars are surrounded by a gamma halo. If the blazar is far enough away, then such halos would appear around each of its images. For close blazars, these images would overlap. It is important that the gravitational lens enhances the radiation flow. Therefore, the a pair halo invisible without lens, can become visible if there is a galaxy (gravitational lens) on the line of sight.


\section{Conclusion}
\label{conclsec}

In this paper, we have considered several effects of gravitational lensing of gamma-ray blazars, taking into account the cascade $e^\pm$ pairs generation in intergalactic magnetic fields. The main goal was to investigate whether gravitational lensing could provide new information about the value of intergalactic magnetic fields or constrain them. Currently, using known cases of gravitational lensing of gamma-ray blazars, it is not yet possible to reliably measure magnetic fields due to unsuccessful lensing parameters, but we hope that new lensed blazars with a suitable configuration will be discovered, for which the method under consideration can be applied.

Unique opportunities for studying gamma-ray lensing and measuring magnetic fields will appear in the future when the angular resolution of gamma-ray telescopes increases by several orders of magnitude, which will allow distinguishing two lensed gamma-ray images and determining the configuration of a gravitational-magnetic lens with high accuracy. Really, as it was shown in the Section~\ref{delay} on the example of the blazar B0218+357, the angular distance between the two images of the blazar is about 0.1". The current and planned in the near future gamma-ray telescopes such as e-ASTROGAM, GECAM, and SVOM have sufficient time resolution and sensitivity to observe a time delay of about 2 days in the strong gravitational lensing of gamma-ray blazars. If new cases of strong lensing are detected, these telescopes will be able to participate in the observations and measure this time delay. However, to observe individual lensed images, an angular resolution of the order of 0.1" is required. Unfortunately, these telescopes do not give such a resolution. Measurement of intergalactic magnetic fields using angular resolution will become possible with the lanch of new gamma-ray telescopes with higher angular resolution.  

If the intergalactic magnetic fields have a relatively large value, $B\geq10^{-16}$~G, then the birth of particles from $e^+e^-$-pairs deviate by large angles, and there will be no isolated trajectories of secondary gamma radiation. Instead, around blazars (not around visible images, but around the objects themselves), so-called halos of pairs (pair halos) are formed. It is possible that gravitational lenses, due to their reinforcing properties, will provide the opportunity to observe such a situation.

The author is grateful to the late V.S. Berezinsky for discussing the method proposed in this article, and to anonymous Reviewers for useful comments.

\end{document}